\documentclass[12pt]{article}
\usepackage{amssymb}
\usepackage{amsmath}

\begin{document}

\noindent

\texttt{The Atlantic Canada General Relativity Regional Meeting}

\texttt{\noindent University of New Brunswick, Fredericton, NB }

\texttt{May 5-7, 2006}

\bigskip

\vspace{3cm}

\begin{center}
{\Huge Averaging Problem in Cosmology\vspace{0.3cm}}

{\Huge and Macroscopic Gravity}
\end{center}

{\large \bigskip }

\begin{center}
{\large Roustam Zalaletdinov}

{\large \bigskip }

{\large \bigskip }

\textsl{Department of Mathematics, Statistics and Computer Science}

\textsl{St Francis Xavier University, Antigonish, Nova Scotia}

\bigskip

{\large \bigskip }\bigskip

\bigskip
\end{center}

\bigskip

\bigskip

\begin{center}
\underline{{\large Contents}}

{\large \bigskip }
\end{center}

{\large \noindent 1. Introduction to General Relativity}

{\large \noindent 2. General Relativity }$\longmapsto ${\large \
Relativistic Cosmology}

{\large \noindent 3. Introduction to Relativistic Cosmology}

{\large \noindent 4. Relativistic Cosmology }$\longmapsto ${\large \
Mathematical Cosmology}

{\large \noindent 5. Averaging Problem in Relativistic Cosmology}

{\large \noindent 6. History of the Averaging Problem}

{\large \noindent 7. Averaging and FLRW Cosmologies}

{\large \noindent 8. Averaging Problem in General Relativity}

{\large \noindent 9. Macroscopic Gravity and General Relativity}

{\large \noindent 10. Macroscopic Gravity and Cosmology}

{\large \noindent 11. The System of Macroscopic Gravity Equations}

{\large \noindent 12. An Exact Cosmological Solution}

{\large \newpage }

\begin{center}
{\LARGE 1. Introduction to General Relativity}
\end{center}

\vspace{1cm}

\noindent {\large $\Rightarrow$ Since the discovery of the General
Relativity Theory in 1915 by A. Einstein it is believed to be a correct
theory of classical gravitational phenomena. }

\vspace{1cm}

\noindent {\large $\Rightarrow$ The starting point in the classical
studies of gravitational phenomena is to determine the so-called line
element }
\begin{equation}
\fbox{$ds^{2}=g_{\alpha \beta }dx^{\alpha }dx^{\beta }$}  \label{line}
\end{equation}
{\large where }$ds^{2}${\large \ is the distance between two neighboring
points }$x^{\alpha }+dx^{\alpha }${\large \ and }$x^{\alpha }${\large , }$
dx^{\alpha }${\large \ is an infinitesimal displacement vector along the
corresponding coordinate lines which are labeled by the indices }$\alpha
,\beta ,\gamma ,...=0,1,2,3${\large \ and }$g_{\alpha \beta }${\large \ is
the metric tensor. The index }$0${\large \ stands usually for one dimension
of time, and indices from }$1${\large \ to }$3${\large \ for 3 space
dimensions. }

\vspace{1cm}

\noindent {\large $\Rightarrow$ The distance }$ds^{2}${\large \ defines
the geometrical properties of a 4-dimensional manifold of time and space,
which is called a space-time. It is known that the geometry of general
relativistic space-times is pseudo-Riemannian.}

\vspace{1cm}

\noindent {\large  $\Rightarrow$  The field equations of General Relativity
are Einstein's equations} {\large \ }
\begin{equation}
R_{\alpha \beta }-\frac{1}{2}g_{\alpha \beta }R=-\kappa T_{\alpha \beta
}\quad \Longleftrightarrow \quad \fbox{\textrm{GEOMETRY=MATTER}}  \label{EE}
\end{equation}
{\large where }$R_{\alpha \beta }${\large \ and }$R${\large \ are the Ricci
tensor and scalar curvature, }$T_{\alpha \beta }${\large \ is an
energy-momentum tensor of matter and radiation, }$\kappa $ {\large is the
Einstein constant, }$\kappa =8\pi G/c^{4}${\large ,\ where }$G${\large \ is
Newton's gravitational constant and }$c${\large \ is the velocity of light.
Einstein's equations are second order partial differential equations for 10
unknown components of the metric tensor }$g_{\alpha \beta }.$

\vspace{1cm}

\noindent {\large  $\Rightarrow$  The equations of motion for matter and
radiation follow from Einstein's equations }
\begin{equation}
T^{\alpha \beta }{}_{;\beta }=0,  \label{em}
\end{equation}
{\large as a result of the pseudo-Riemannian geometry of space-time. These
equations are employed in order to interpret the local (e.g. solar system)
and large scale (e.g. cosmological) observations, on the one hand, and to
construct mathematical models in order to predict the evolution of
gravitational phenomena, on the other.}

\newpage

\begin{center}
{\LARGE 2. General Relativity }$\longmapsto ${\LARGE \ Relativistic
Cosmology }
\end{center}

\vspace{1cm}

\noindent {\large  $\Rightarrow$  From a theoretical point of view, the main
idea is to treat the equations of motion (\ref{em}) as a correspondence
rule. Whereby given the form of the energy-momentum tensor }$T_{\alpha \beta
}$ {\large and assuming a structure of the corresponding metric tensor (\ref{line}),
the space-time geometry can be specified by solving Einstein's
equations (\ref{EE}). The resulting predictions of the theory are then to be
compared with real observations, in order to ascertain its viability as a
theory of gravity on all relevant scales, including the cosmological ones. }

\vspace{1cm}

\begin{center}
\fbox{{\large General Relativity it is considered valid on all classical
physical scales}}
\end{center}

\vspace{1cm}

\noindent {\large  $\Rightarrow$  It is well known that general relativity
has been extremely successful in accounting for local observations,
including the classical tests in the Solar System and the observations of
binaries. Here, however, it is instructive to contrast this with the status
of General Relativity as a theory of gravity on cosmological scales, for
which there is much less detailed direct and observational evidence. The
main reason for this is not just the usual \textquotedblleft
uncertainty\textquotedblright\ brought about by the errors of our
measurements which are bound to be present in all cosmological observations.
It relates to the very nature of real cosmological observations and the
difficulties that arise when attempts are made to construct an appropriate
theoretical framework for cosmological models in General Relativity. }

\vspace{1cm}

\noindent {\large  $\Rightarrow$  Our understanding of general relativistic
Cosmology, that is, the large-scale structure of our Universe, has been
unchanged now for decades since the discovery (Friedmann - 1922, Lema\^{\i}
tre - 1927, Robertson -1935, Walker -1935) of the Friedmann-Lema\^{\i}
tre-Robertson-Walker (FLRW) cosmological solutions to Einstein's equations.
They assume that in some statistical sense, or averaged over large volumes,
the Universe is spatially homogeneous and isotropic. Such a state is
described by the Robertson-Walker line element }
\begin{equation}
ds_{\mathrm{FLRW}}^{2}=-dt^{2}+a^{2}(t)[\frac{dr^{2}}{(1-kr^{2})}
+r^{2}(d\theta ^{2}+\sin ^{2}\theta d\phi ^{2})]  \label{RW}
\end{equation}
{\large where} $a(t)${\large \ is the expansion factor of Universe and the
parameter }$k${\large \ is the curvature parameter which is equal to }$+1$
{\large \ for a closed model of Universe, }$-1$ {\large for an open model
and }$0${\large \ for a flat model where 3-dimensional homogeneous and
isotropic spaces are 3-spheres, hyperbolic or flat 3-spaces,
correspondingly. Hereafter the unit system }$c=1$ {\large is used. }

\vspace{1cm}

\newpage

\begin{center}
{\LARGE 3. Introduction to Relativistic Cosmology}
\end{center}

\vspace{1cm}

\noindent {\large  $\Rightarrow$  Einstein's equations (\ref{EE}) for the
line element (\ref{RW}) called Friedmann's equation read}
\begin{equation}
\left( \frac{1}{a}\frac{da}{dt}\right) ^{2}=\frac{\kappa \rho }{3}-\frac{k}{
a^{2}}  \label{F}
\end{equation}
{\large where }$\rho ${\large \ is the matter density. Friedmann's equation
describes the evolution of the Universe as either an expanding, collapsing
or being spatially flat space-time. }

\vspace{1cm}

\noindent {\large  $\Rightarrow$  The factor }$H=\overset{\cdot }{a}/a$
{\large \ called the Hubble parameter measures the rate }$v$ {\large of
expansion or collapse as }$v=HR$, $R=a(t)r$. {\large The curvature parameter
}$k${\large \ can be expressed through the current value of Hubble parameter
}$H_{0}=100$(\textrm{km/s} \textit{Mps})$h_{0}$,{\large \ }$1/2<h_{0}<1$,
{\large and the parameter }$\Omega _{0}=\rho _{0}/\rho _{c}${\large \ which
measures the ratio of the current value of matter density }$\rho _{0}$
{\large \ to the critical one }$\rho _{c}$, {\large \ }
\begin{equation}
\fbox{$k=a_{0}^{2}H_{0}^{2}\left( \Omega _{0}-1\right) $}  \label{k}
\end{equation}
{\large This fundamental formula relates the physics of the Universe with
its global structure, topology, its matter density and the character of its
evolution. }

\vspace{1cm}

\noindent {\large  $\Rightarrow$  The FLRW models serve as the main
theoretical laboratory for prediction, development and verification of our
astrophysical and cosmological experiments and data. Up to the end of XXth
century the whole body of cosmological evidence had been showing in favor of
an open model when }$\Omega _{0}=\rho _{0}/\rho _{c}<1$.

\vspace{1cm}

\noindent {\large  $\Rightarrow$  Despite the enormous progress made in
understanding the Universe structure on the basis of FLRW\ models, there is,
however, a number of theoretical and experimental issues which are in
contradiction with them. }

\vspace{1cm}

\noindent {\large  $\Rightarrow$  Age of Universe: Calculation of the age of
Universe for a matter-dominated era for a flat model gives the value}
\begin{equation}
t_{0}=\frac{2}{3H_{0}}=\frac{20}{3}h_{0}^{-1}\mathrm{Gyr}  \label{age}
\end{equation}
{\large that is }$7\times 10^{9}${\large \ \textrm{yr} }$<t_{0}<1.3\times
10^{10}${\large \ \textrm{yr}}, {\large which far beyond the observational
limits. For open and closed models the situation is even worse.}

\vspace{1cm}

\bigskip

\begin{center}
{\LARGE 3. Introduction to Relativistic Cosmology - 2}
\end{center}

\vspace{1cm}

\noindent {\large  $\Rightarrow$  Dark Matter of Universe: The observational
evaluation of the luminous (baryonic) matter of the Universe shows its
presence in the amount of}
\begin{equation}
\Omega _{\mathrm{bar}}=\rho _{\mathrm{bar}}/\rho _{c}=0.03
\label{bar-matter}
\end{equation}
{\large However, observations of the dynamics of galaxies and clusters of
galaxies have revealed the dynamical effects due to some unseen, "dark"
matter which interacts only gravitationally. Its nature is unknown, but it
cannot be baryonic, that is, composed from known so far particles which form
the luminous (baryonic) matter (hydrogen, helium, lithium, boron, etc.). The
evaluation of the amount of this mysterious dark matter gives }
\begin{equation}
\Omega _{\mathrm{dark}}=\rho _{\mathrm{dark}}/\rho _{c}=0.3
\label{dark-matter}
\end{equation}
{\large which is 10 times more than usual forms of matter and radiation. }

\vspace{1cm}

\noindent {\large  $\Rightarrow$  Acceleration of Universe: In 1998 the
measurements of type Ia supernova light curves, in particular SN 1997ff,
have provided clear evidence that the Universe is speeding up, not slowing
down, as it has been and should be expected from FLRW models. This is rather
surprising, unexpected discovery led to an assumption of the existence of a
cosmological agent called the dark energy which causes this acceleration.
Its nature is not known as yet, but, apparently, it emits no light, i.e. it
is "dark", and it must have its pressure to be negative and comparable in
magnitude with its energy density, i.e. it is not any kind of "matter". Dark
energy can only interact gravitationally. The evaluation of the contribution
due to this mysterious dark energy gives }
\begin{equation}
\Omega _{\Lambda }=\rho _{\Lambda }/\rho _{c}=0.7  \label{dark-energy}
\end{equation}
{\large which is about 2/3 of all energy-matter content the Universe
provided }$\Omega _{0}=1${\large \ or close to 1. The current hypotheses
about dark energy mostly assume that it originates from a cosmological
constant }$\Lambda $ {\large which can be incorporated into Einstein's
equations. }

\vspace{1cm}

\noindent {\large  $\Rightarrow$  Flatness of Universe: Observations and
measurements of the properties of cosmic microwave background (CMB),
especially, the anisotropy of CMB, carried out for a few decades, and
recently in 2000-2002 with the much improved resolution, imply that the
parameter }$\Omega _{0}=\rho _{0}/\rho _{c}${\large \ which measures the
ratio of the current value of matter density }$\rho _{0}${\large \ to the
critical one }$\rho _{c}${\large , is }
\begin{equation}
\Omega _{0}=1\pm 0.04.  \label{omega}
\end{equation}
{\large Thus, }$k=0${\large \ (\ref{k}) and our Universe appears to be
spatially flat. }

\vspace{1cm}

\bigskip

\begin{center}
{\LARGE 3. Introduction to Relativistic Cosmology - 3}
\end{center}

\vspace{1cm}

\noindent {\large  $\Rightarrow$ ... And having almost all its
energy-matter content in totally "dark", unseen forms: }
\begin{equation}
\Omega _{\mathrm{dark}}+\Omega _{\Lambda }+\Omega _{\mathrm{bar}}=1\gg
\Omega _{\mathrm{bar}}.  \label{omega=1}
\end{equation}

{\large \ \newpage }

\begin{center}
\bigskip {\LARGE 4. Relativistic Cosmology }$\longmapsto ${\LARGE \
Mathematical Cosmology}
\end{center}

\vspace{1cm}

\noindent {\large  $\Rightarrow$  This value of the age of Universe is in
disagreement with the most likely values for globular cluster ages. The
situation with the dark matter and dark energy even worse since even no
reliable physical hypotheses exist and can be proposed for now. It can be
improved for the latter to some extent by taking into account a positive
cosmological constant, but its introduction into Einstein's equations
(\ref{EE}) does change their structure. But, that opens other problems with the
nature and properties of such a new physical constant. }

\vspace{1cm}

\noindent {\large  $\Rightarrow$  These problems refers to the so-called
Standard model consistency which calls for observational and theoretical
tests of consistency of the standard FLRW model. }

\vspace{1cm}

\noindent {\large  $\Rightarrow$  Relativistic Cosmology is based on the
Cosmological Principle: }

\begin{center}
\textbf{The large-scale structure of the Universe (at least in its
observable part) is }

\textbf{well-described by expanding homogeneous and isotropic FLRW\ geometry}
\end{center}

\noindent {\large  $\Rightarrow$ The main goal of Relativistic Cosmology is
to find out whether or not the Cosmological Principle holds, that is, if it
does provide realistic observational physics of the evolving Universe in an
adequate geometrical space-time picture. }

\begin{center}
\fbox{\textbf{Relativistic Cosmology = Observational Cosmology + Theoretical
Cosmology}}

\bigskip
\end{center}

\noindent {\large  $\Rightarrow$  Observational Cosmology is determining the
space-time geometry and physics of the Universe from the observational data:}

\vspace{0.5cm}

$\blacktriangleright $ observational constraints: we essentially see the
Universe from one space-time point down one null cone

$\blacktriangleright $ observational cosmology programme: how to determine
the space-time geometry from null cone data

$\blacktriangleright $ proof of homogeneity: we observe a high degree of
isotropy, but what about spatial geometry?

$\blacktriangleright $ standard model parameters: $H_{0}$, $\Omega _{0}$, $
\rho _{c}...$,

$\blacktriangleright $ standard FLRW model consistency: age of Universe,
dark matter, dark energy, etc.

\vspace{1cm}

\noindent {\large  $\Rightarrow$  Theoretical Cosmology called often
Mathematical Cosmology is determining the space-time geometry and physics of
the Universe from known physical theories to fit the observational data
and/or proposing new theoretical models capable of more realistic
description.}

\newpage

\begin{center}
{\LARGE 5. Averaging Problem in Relativistic Cosmology}

\vspace{1cm}
\end{center}

\noindent {\large  $\Rightarrow$  Mathematical Cosmology: }

\vspace{0.5cm}

$\bullet $ Cosmological Dynamics - the nature of the dynamical Universe
evolution (description, equations, consistency, acceleration,
isotropization, singularities, etc.),

$\bullet $ Realistic Models - the dynamical and evolutional content of
relation of local lumpiness to the overall smooth cosmos (fitting and
averaging, creation of structures, horizons, arrow of time, etc.), dark
matter models, dark energy models.

$\bullet $ Inflationary and Cosmological Creation - the dynamical picture of
growth of local structures into the large-scale structures of physical
Universe, etc.

\vspace{1cm}

\begin{center}
\fbox{{\large The real Universe is lumpy, not smooth!}}

\vspace{1cm}
\end{center}

\noindent {\large  $\Rightarrow$  Our Universe seems to be isotropic and
homogeneous at very large scales as supported by the observed isotropy of
the cosmic microwave background. The present state of the actual Universe
is, however, neither homogeneous nor isotropic on scales }$\lesssim 100$
\textit{Mps. }

\vspace{1cm}

\noindent {\large  $\Rightarrow$  Since the FLRW\ models, and more
generally, Relativistic Cosmology of the Universe relies on the Cosmological
Principle, the problem of dynamical justification of the large-scale smooth
structure, given a lumpy inhomogeneous dynamics on smaller scales, takes the
central place in Relativistic Cosmology (Ellis - 1984). }

\vspace{1cm}

\begin{center}
\fbox{{\large The Averaging Problem in Cosmology}}
\end{center}

\bigskip

\begin{center}
{\large Does an inhomogeneous Universe evolve on average like a homogeneous
one?}

$\blacktriangledown $

{\large Why is the large-scale volume behavior of the lumpy Universe }

{\large the same as that predicted by the smoothed out FLRW\ models?}

$\blacktriangledown $

{\large Is a homogeneous solution of Einstein's equations the same }

{\large as an averaged solution for a inhomogeneous space-time?}

$\blacktriangledown $

{\large Are Einstein's equations \textquotedblleft the
same\textquotedblright\ when applied on different physical scales?}
\end{center}

\newpage

\begin{center}
{\LARGE 6. History of Averaging Problem}
\end{center}

\vspace{1cm}

\noindent {\large  $\Rightarrow$  \underline{The main difficulties in doing
averaging procedures in General Relativity}:}

\vspace{0.5cm}

\noindent {\large 1. Covariant volume averages (averages over compact
subspaces) are difficult to define on a pseudo-Riemannian space-time because
the integration of tensors is not generally defined on curved manifolds. }

\vspace{0.5cm}

\noindent {\large 2. Statistical or ensemble averages also meet difficulties
because they require a curved space-time manifold to be split into time and
3-space which cannot be accomplished generally.}

\vspace{0.5cm}

\noindent {\large 3. General Relativity is a nonlinear theory for unknown
components of metric tensor }$g_{\alpha \beta }${\large \ and in general }
\begin{equation}
\left\langle \left( R_{\alpha \beta }-\frac{1}{2}g_{\alpha \beta }R\right)
[g_{\rho \sigma }]\right\rangle \neq \left( \langle R_{\alpha \beta }\rangle
-\frac{1}{2}\langle g_{\alpha \beta }R\rangle \right) [\langle g_{\rho
\sigma }\rangle ]\   \label{noncommute}
\end{equation}
{\large where }$\langle \cdot \rangle $ {\large stands for an average
operator provided it is defined. One must split out the products}
\begin{equation}
\langle g_{\alpha \beta }R\rangle =\langle g_{\alpha \beta }\rangle \langle
R\rangle +C(g_{\alpha \beta },R)  \label{correlation}
\end{equation}
{\large where} $C(g_{\alpha \beta },R)$ {\large is a kind of a correlation
tensor.}

\bigskip

\begin{center}
\fbox{{\large Averaging does modify Einstein's equations by an effective
energy term}}

\bigskip
\end{center}

\noindent {\large  $\Rightarrow$  The overwhelming majority of approaches
(Shirokov \& Fisher - 1962, Isaacson - 1968, Szekeres - 1971, Noonan - 1984,
Boersma -1998, Zotov \& Stoeger, 1992, 1995, Stoeger et al - 1999, see
review in Krasi\'{n}ski - 1997) made use of }

\vspace{0.5cm}

$\blacksquare $ a {\large perturbation theory where the line element has the
form }
\begin{equation}
ds_{\mathrm{pert}}^{2}=(g_{\alpha \beta }^{(B)}+h_{\alpha \beta })dx^{\alpha
}dx^{\beta }  \label{perturb}
\end{equation}
{\large where }$g_{\alpha \beta }^{(B)}${\large \ is the background metric
tensor for a smooth background space-time and} $h_{\alpha \beta }${\large \
are the perturbation functions small compared with }$g_{\alpha \beta }^{(B)}$

\bigskip

$\blacksquare $ {\large various non-covariant space and space-time volume
averaging procedures}

\bigskip

\noindent {\large to arrive at the modified Einstein's equations }
\begin{equation}
R_{\alpha \beta }^{(B)}-\frac{1}{2}g_{\alpha \beta }^{(B)}R^{(B)}=-\kappa
T_{\alpha \beta }+C_{\alpha \beta }  \label{EEmod}
\end{equation}
{\large where the correlation tensor }$C_{\alpha \beta }$ {\large is defined
as an average of quadratic terms in perturbation functions} $h_{\alpha \beta
}${\large \ and it is the effective energy in physical meaning.}

\newpage

\begin{center}
{\LARGE 7. Averaging and FLRW Cosmologies}

\vspace{1cm}
\end{center}

\noindent {\large  $\Rightarrow$  Inhomogeneous cosmological models which
are homogeneous and isotropic on average: }

\bigskip

$\blacklozenge $ {\large 3+1 cosmological space-time splitting with space
volume averages (Kasai - 1992, 1993, Futamase - 1996, Buchert - 1999-2005). }

\bigskip

$\blacklozenge $ {\large perturbed FLRW Universe with space averages
(Futamase - 1988, 1991, 1993, Bildhauser \& Futamase -1991).}

\bigskip

\noindent {\large  $\Rightarrow$  Backreaction in Cosmology for a perturbed
FLRW Universe with spatial volume averages (Bildhauser - 1990) and applying
the renormalization group methods (Nambu - 2000).}

\bigskip

\noindent {\large  $\Rightarrow$  Direct averaging of Friedmann's equation
with space volume averages (Russ, Soffel, Kasai, B\"{o}rner - 1997).}

\bigskip

\noindent {\large  $\Rightarrow$  Averaging Newtonian cosmologies with
spatial volume averages (Buchert \& Ehlers - 1997, Iguchi, Hosoya \& Koike
-1998).}

\bigskip

\begin{center}
\fbox{{\large The size of the averaging space regions is tacitly assumed to
be }$\simeq ${\large \ }$100${\large \ }\textit{Mps}}

\vspace{1cm}
\end{center}

\noindent {\large  $\Rightarrow$  Modified Friedmann's equation by the
appearance of effective average energy density} $\rho _{\mathrm{cor}}\sim
\langle (\partial h)^{2}\rangle ${\large \ }
\begin{equation}
\left( \frac{1}{a}\frac{da}{dt}\right) ^{2}=\frac{\kappa }{3}\left( \rho
+\rho _{\mathrm{cor}}\right) -\frac{k}{a^{2}}.  \label{Fmod}
\end{equation}

\bigskip

\begin{center}
\fbox{\textbf{An inhomogeneous Universe which is FLRW on average evolves
differently!}}

\vspace{1cm}
\end{center}

$\bigstar ${\large \ The evolutional effect of the effective average energy
density} $\rho _{\mathrm{cor}}$ {\large due to the self-interaction of
gravitational inhomogeneities propagating on a FLRW\ homogeneous and
isotropic cosmological space-time is able to change the age of the flat
Universe up to larger than }$17$ \textrm{Gyr }{\large as compared with }$7$
{\large \ }\textrm{Gyr}{\large \ }$<t_{0}<13${\large \ }\textrm{Gyr}{\large
\textrm{\ }in the standard FLRW\ picture - without any other additional
hypothesis like the presence of cosmological constant or other cosmological
fields.}

\bigskip

$\bigstar ${\large \ A realistic cosmological model which reflects the real
inhomogeneous structure of our Universe on smaller scales than the
cosmological one does promise to fit our observational data. }\newpage

\begin{center}
{\LARGE 8. Averaging Problem in General Relativity}
\end{center}

\vspace{1cm}

\noindent {\large  $\Rightarrow$  The standard approach in Relativistic
Cosmology based on the Cosmological Principle:}

\bigskip

\noindent $\blacktriangledown $ \underline{{\large Assumption 1}}{\large :
In the real complicated \textquotedblright lumpy\textquotedblright\ Universe
with a discrete matter distribution on each physically distinct scale (of
stars, galaxies, clusters of galaxies, etc.), the stress-energy tensor} $
T_{\alpha \beta }=T_{\alpha \beta }^{\mathrm{(discrete)}}$ {\large can be
adequately approximated by a \textquotedblright smoothed\textquotedblright ,
or hydrodynamic, stress-energy tensor }
\begin{equation}
T_{\alpha \beta }=T_{\alpha \beta }^{\mathrm{(hydro)}}=\langle T_{\alpha
\beta }^{\mathrm{(discrete)}}\rangle ,  \label{<T>}
\end{equation}
{\large usually taken to be representable by a simple perfect fluid.}

\bigskip

\noindent $\blacktriangledown $\underline{ {\large Assumption 2}}{\large :}
{\large As} $T_{\alpha \beta }^{\mathrm{(discrete)}}\rightarrow T_{\alpha
\beta }^{\mathrm{(hydro)}}$ {\large on the right-hand side of Einstein's
equations (\ref{EE}), the structure of the field operator in the left-hand
side is kept unchanged under such a change and therefore the appropriate
field equations for describing the matter distribution }$T_{ab}^{\mathrm{
(hydro)}}$ {\large are taken in the form}
\begin{equation}
\overline{R}_{\alpha \beta }-\frac{1}{2}G_{\alpha \beta }\overline{R}
=-\kappa T_{\alpha \beta }^{\mathrm{(hydro)}}  \label{EEst}
\end{equation}
{\large where} $G_{ab}$ {\large , }$\overline{R}_{ab}${\large \ and }$
\overline{R}$ {\large are the metric, Ricci tensors and scalar curvature of
a pseudo-Riemannian space-time geometry describing now the corresponding
geometry created by the smoothed }$T_{\alpha \beta }^{\mathrm{(hydro)}}$
{\large . } {\large The corresponding equations of motion following from (
\ref{EEst}) are}
\begin{equation}
T_{;\beta }^{\alpha \beta \mathrm{(hydro)}{}}={}0  \label{em-st}
\end{equation}

\begin{center}
{\large Is there a firm physical and geometrical evidence that} {\large
(\ref{<T>}) does imply }$R_{\alpha \beta }-\frac{1}{2}g_{\alpha \beta
}R\rightarrow \overline{R}_{\alpha \beta }-\frac{1}{2}G_{\alpha \beta }
\overline{R}=\langle R_{\alpha \beta }-\frac{1}{2}g_{\alpha \beta }R\rangle $
?

\vspace{1cm}

$\Downarrow $

\vspace{1cm}

\fbox{\textbf{On what scale do Einstein's equations hold?}}

\bigskip
\end{center}

$\bigstar $ {\large General Relativity as a classical theory of gravity is
likely to be a microscopic theory (Tavakol \& Zalaletdinov - 1998) which is
physically adequate and realistic for treatment of the gravitational field
created by a point-like discrete matter distribution.}

\bigskip

$\bigstar $ {\large The Averaging Problem of the fundamental importance
arises in the applications of Einstein's equations in Cosmology and other
settings where adequate matter distributions are smooth and continuous. }

\newpage

\begin{center}
{\LARGE 9. Macroscopic Gravity and General Relativity}
\end{center}

\vspace{1cm}

\noindent {\large  $\Rightarrow$  Macroscopic gravity is a non-perturbative
geometrical approach (Zalaletdinov - 1992-2005) to resolve the Averaging
Problem: a reformulation in a broader context as the problem of macroscopic
description of gravitation }

\bigskip

$\blacktriangle ${\large \ Classical physical phenomena possess two levels
of description (Lorentz, 1897, 1916):}

\bigskip

\begin{center}
\fbox{\textbf{The microscopic description }$\mathbf{\Longleftrightarrow }$
\textbf{\ The discrete matter model}}
\end{center}

\vspace{0.05cm}

\begin{center}
$\downarrow $ \texttt{by a suitable averaging procedure} $\downarrow $

\bigskip

\fbox{\textbf{The macroscopic description }$\mathbf{\Longleftrightarrow }$
\textbf{\ The continuous matter model}}
\end{center}

\bigskip

\noindent \underline{{\large Lorentz 'theory of electrons}} \hspace{4.5cm}
{\large \ }\underline{{\large Maxwell's electrodynamics}}

\begin{equation*}
F_{,\nu }^{\mu \nu }=\frac{4\pi }{c}j^{\mu }=4\pi
\sum\limits_{i}q_{i}u^{\mu }(t_{i})\hspace{1cm}\rightarrow \langle \mathrm{
averaging}\rangle \rightarrow \hspace{1cm}H^{\mu \nu }{}_{,\nu }=\frac{4\pi
}{c}\langle j\rangle ^{\mu }=\frac{4\pi }{c}(J^{\mu }-cP_{,\nu }^{\mu \nu })
\end{equation*}

\begin{equation*}
F_{[\alpha \beta ,\gamma ]}=0\hspace{2.5cm}\rightarrow \langle \mathrm{
averaging}\rangle \rightarrow \hspace{2cm}\langle F\rangle _{\lbrack \alpha
\beta ,\gamma ]}=0,~H^{\mu \nu }=\langle F\rangle ^{\mu \nu }+4\pi P^{\mu
\nu }
\end{equation*}

\vspace{1cm}

\noindent {\large  $\Rightarrow$  The case of General Relativity is much
more complicated:}

\bigskip

\fbox{\textsc{averages}}{\large \ Definition of covariant space, space-time
volume or statistical averages on pseudo-Riemannian space-times}

\bigskip

\fbox{\textsc{geometry}}{\large \ The pseudo-Riemannian geometry of
space-time, the nonlinear structure of the field operator of Einstein's
equations and necessity to deal with gravitational field correlators }

\bigskip

\fbox{\textsc{avmatter}}{\large \ The problem of construction of models of
smoothed, continuously distributed self-gravitating media }$T_{\alpha \beta
}^{\mathrm{(hydro)}}=\langle T_{\alpha \beta }^{\mathrm{(discrete)}}\rangle
. $

\bigskip

$\bigstar $ {\large Einstein's equations themselves are not sufficient to be
consistently averaged out}

\begin{equation*}
\langle R_{\alpha \beta }\rangle -\frac{1}{2}\langle g_{\alpha \beta }g^{\mu
\nu }R_{\mu \nu }\rangle =-\kappa \langle T_{\alpha \beta }^{\mathrm{(micro)}
}\rangle \Rightarrow \langle R_{\alpha \beta }\rangle -\frac{1}{2}\langle
g_{\alpha \beta }\rangle \langle g^{\mu \nu }\rangle \langle R_{\mu \nu
}\rangle +C_{\alpha \beta }=-\kappa \langle T_{\alpha \beta }^{\mathrm{
(micro)}}\rangle \vspace{0.5cm}
\end{equation*}
{\large They become a definition of the correlation function }$C_{\alpha
\beta }$ {\large unless this object is defined from outside the averaged
Einstein's equations. }

\begin{center}
\bigskip \newpage {\LARGE 10. Macroscopic Gravity and Cosmology}
\end{center}

\vspace{1cm}

\noindent {\large  $\Rightarrow$ }\fbox{\textsc{averages}}{\large \ A
covariant space-time volume averaging procedure for tensor fields on
pseudo-Riemannian space-time has been developed by defining particular
averaging kernels which provide well-defined analytic properties of average
field. Such kernels have been shown to always exist.}

\bigskip

\noindent {\large  $\Rightarrow$  }\fbox{\textsc{geometry}}{\large \ The
structure of the macroscopic (averaged) space-time geometry has been found
by developing a procedure of a volume averaging out of Cartan's structure
equations describing the pseudo-Riemannian geometry of space-time. The
definition of all involved gravitational field (geometrical) correlators has
been given.}

\bigskip

\noindent {\large  $\Rightarrow$  }\fbox{\textsc{avmatter}}{\large \ The
problem of construction of models of smoothed, continuously distributed
self-gravitating media }$T_{\alpha \beta }^{\mathrm{(hydro)}}=\langle
T_{\alpha \beta }^{\mathrm{(discrete)}}\rangle $ {\large has been analyzed
recently (Montani, Ruffini \& Zalaletdinov - 2000, 2003, 2004). }

\bigskip

\begin{center}
\fbox{{\large The macroscopic space-time geometry is non-Riemannian}}

\bigskip
\end{center}

$\bigstar $ {\large The field equations of Macroscopic Gravity are the
averaged Einstein's equations}
\begin{equation}
\bar{g}^{\alpha \epsilon }M_{\epsilon \beta }-\frac{1}{2}\delta _{\beta
}^{\alpha }\bar{g}^{\mu \nu }M_{\mu \nu }=-\kappa \langle T_{\beta }^{\alpha
\mathrm{(micro)}}\rangle +(Z^{\alpha }{}_{\mu \nu \beta }-\frac{1}{2}\delta
_{\beta }^{\alpha }Q_{\mu \nu })\bar{g}^{\mu \nu }  \label{MG}
\end{equation}
{\large where} $\bar{g}^{\mu \nu }$ {\large is the averaged metric,} $M_{\mu
\nu }$ {\large is the Ricci tensor of the Riemannian curvature} $M^{\alpha
}{}_{\beta \gamma \delta }$ {\large which stands for the} {\large induction
tensor,} $Z^{\alpha }{}_{\mu \nu \beta }-\frac{1}{2}\delta _{\beta }^{\alpha
}Q_{\mu \nu }$ {\large is the correlation tensor constructed from the
correlation connection tensor,}
\begin{equation}
Z^{\alpha }{}_{\beta \gamma }{}^{\mu }{}_{\nu \sigma }\equiv Z^{\alpha
}{}_{\beta \lbrack \gamma }{}^{\mu }{}_{\underline{\nu }\sigma ]}=\langle
\Gamma ^{\alpha }{}_{\beta \lbrack \gamma }{}\Gamma ^{\mu }{}_{\underline{
\nu }\sigma ]}\rangle -\langle \Gamma ^{\alpha }{}_{\beta \lbrack \gamma
}{}\rangle \langle \Gamma ^{\mu }{}_{\underline{\nu }\sigma ]}\rangle
\label{Z6}
\end{equation}
.

\bigskip

$\bigstar $ {\large There are separate equations for correlation tensors. }

\bigskip

$\bigstar $ {\large There is also a non-Riemannian curvature tensor which
stands for average field tensor and it satisfies its own equations. }

\begin{center}
\bigskip \newpage {\LARGE 10. Macroscopic Gravity and Cosmology - 2}
\end{center}

\vspace{1cm}

\noindent {\large   $\Rightarrow$ A study of the structure of the field
equations of macroscopic gravity enables one to answer a fundamental
question of Relativistic Cosmology about the physical meaning and the range
of applicability of Einstein's equations with a continuous (smoothed) matter
source:}

\bigskip

\begin{center}
\textbf{The averaged Einstein's equations become Einstein's equations}
\begin{equation}
M_{\alpha \beta }-\frac{1}{2}G_{\alpha \beta }G^{\mu \nu }M_{\mu \nu
}=-\kappa T_{\alpha \beta }^{\mathrm{(hydro)}}  \label{MG->EE}
\end{equation}
\textbf{for the macroscopic metric} $G_{\alpha \beta }$ \textbf{with a
smoothed stress-energy tensor if }

\vspace{1cm}

\fbox{\textbf{all correlations functions vanish}}

\bigskip
\end{center}

\newpage

\begin{center}
{\LARGE 11. The System of Macroscopic Gravity Equations}
\end{center}

\vspace{1cm}

\noindent {\large  $\Rightarrow$  The Riemannian Sector:}

{\large \medskip \bigskip }

{\large (1.A) Define a line element for the macroscopic geometry}
\begin{equation}
ds^{2}=G_{\alpha \beta }dx^{\alpha }dx^{\beta }  \label{@ds^2}
\end{equation}
{\large in terms of the macroscopic metric tensor }$G_{\alpha \beta }$
{\large \ and calculate the Levi-Civita connection coefficients }$\overline{
\mathcal{F}}^{\alpha }{}_{\beta \gamma } \equiv \langle \Gamma ^{\alpha }{}_{\beta \gamma}{}\rangle $
\begin{equation}
\overline{\mathcal{F}}^{\alpha }{}_{\beta \gamma }=\frac{1}{2}G^{\alpha
\epsilon }(G_{\beta \epsilon ,\gamma }+G_{\gamma \epsilon ,\beta }-G_{\beta
\gamma ,\epsilon }).  \label{@Fb=dG}
\end{equation}
{\large and the Riemannian curvature tensor }$M^{\alpha }{}_{\beta \rho
\sigma }${\large \ }
\begin{equation}
M^{\alpha }{}_{\beta \gamma \delta }=2\overline{\mathcal{F}}^{\alpha
}{}_{\beta \lbrack \delta ,\gamma ]}+2\overline{\mathcal{F}}^{\alpha
}{}_{\epsilon \lbrack \gamma }\overline{\mathcal{F}}^{\epsilon }{}_{
\underline{\beta }\delta ]}  \label{@M=DFb}
\end{equation}
{\large in terms of unknown metric functions.}

{\large \medskip \bigskip }

{\large (1.B) Determine the linearly independent components of the
connection correlation tensor }$Z^{\alpha }{}_{\beta \gamma }{}^{\mu
}{}_{\nu \sigma }${\large \ satisfying the algebraic conditions:}

{\large \medskip \bigskip }

{\large (i) the antisymmetry in the third and sixth indices}
\begin{equation}
Z^{\alpha }{}_{\beta \gamma }{}^{\mu }{}_{\nu \sigma }=-Z^{\alpha }{}_{\beta
\sigma }{}^{\mu }{}_{\nu \gamma },  \label{Z+Z=0}
\end{equation}

{\large (ii) the antisymmetry in interchange of the pair of the first
contravariant and the second covariant indices }$\binom{\alpha }{\beta }$
{\large \ with the pair of the second contravariant and the third covariant
indices }$\binom{\mu }{\nu }${\large \ }
\begin{equation}
Z^{\alpha }{}_{\beta \gamma }{}^{\mu }{}_{\nu \sigma }=-Z^{\mu }{}_{\nu
\gamma }{}^{\alpha }{}_{\beta \sigma },  \label{Z2+Z2=0}
\end{equation}

{\large (iii) the algebraic cyclic identities}
\begin{equation}
Z^{\alpha }{}_{\beta [\gamma }{}^{\mu }{}_{\nu \sigma] }=0,  \label{Z^dx=0}
\end{equation}

{\large (iv) the equi-affinity property}
\begin{equation}
Z^{\epsilon }{}_{\epsilon \gamma }{}^{\mu }{}_{\nu \sigma }=0.  \label{TrZ=0}
\end{equation}
{\large The algebraic symmetries (\ref{Z+Z=0}) and (\ref{Z2+Z2=0}) result in}

{\large (v) the symmetry of interchange of the index triples }$\left(
^{\alpha }{}_{\beta \gamma }\right) ${\large \ and }$\left( ^{\mu }{}_{\nu
\sigma }\right) ${\large \ }
\begin{equation}
Z^{\alpha }{}_{\beta \gamma }{}^{\mu }{}_{\nu \sigma }=Z^{\mu }{}_{\nu
\sigma }{}^{\alpha }{}_{\beta \gamma }.  \label{Z3-Z3=0}
\end{equation}
{\large Any algebraic symmetry from (i), (ii) and (v) follows from two other
symmetry properties. There are generally 396 independent components. }

\begin{center}
{\large \bigskip \newpage }{\LARGE 11. The System of Macroscopic Gravity
Equations - 2}
\end{center}

{\LARGE \vspace{1cm}}

{\large (1.C) Make an ansatz about the functional form of the components of
the correlation tensor }$Z^{\alpha }{}_{\beta \gamma }{}^{\mu }{}_{\nu
\sigma }${\large \ on the basis of symmetries and physical conditions of the
macroscopic geometry. }

{\large \medskip \bigskip }

{\large (1.D) solve the integrability conditions for equations (\ref{@DZ=0})
}
\begin{equation}
Z^{\epsilon }{}_{\beta \lbrack \gamma }{}^{\mu }{}_{\underline{\nu }\sigma
}M^{\alpha }{}_{\underline{\epsilon }\lambda \rho ]}-Z^{\alpha }{}_{\epsilon
\lbrack \gamma }{}^{\mu }{}_{\underline{\nu }\sigma }M^{\epsilon }{}_{
\underline{\beta }\lambda \rho ]}+Z^{\alpha }{}_{\beta \lbrack \gamma
}{}^{\epsilon }{}_{\underline{\nu }\sigma }M^{\mu }{}_{\underline{\epsilon }
\lambda \rho ]}-Z^{\alpha }{}_{\beta \lbrack \gamma }{}^{\mu }{}_{\underline{
\epsilon }\sigma }M^{\epsilon }{}_{\underline{\nu }\lambda \rho ]}=0.
\label{@ic:DZ=0}
\end{equation}
{\large to find some independent components of }$Z^{\alpha }{}_{\beta \gamma
}{}^{\mu }{}_{\nu \sigma }${\large \ expressed in terms of unknown metric
functions.}

{\large \medskip \bigskip }

{\large (1.E) solve the system of differential equations for the connection
correlation tensor }$Z^{\alpha }{}_{\beta \gamma }{}^{\mu }{}_{\nu \sigma }$
{\large \ }
\begin{equation}
Z^{\alpha }{}_{\beta \lbrack \gamma }{}^{\mu }{}_{\underline{\nu }\sigma
\parallel \lambda ]}=0,  \label{@DZ=0}
\end{equation}
{\large to find the rest of the components of }$Z^{\alpha }{}_{\beta \gamma
}{}^{\mu }{}_{\nu \sigma }${\large , including }$Z^{\alpha }{}_{(\mu \nu
)\beta }${\large , in terms of unknown metric functions.}

{\large \medskip \bigskip }

{\large The steps (1.A)-(1.E) solve the Riemannian Sector to find the
Levi-Civita connection coefficients }$\overline{\mathcal{F}}^{\alpha
}{}_{\beta \gamma }${\large , the Riemannian curvature tensor }$M^{\alpha
}{}_{\beta \rho \sigma }${\large , a part of components of the connection
correlation tensor }$Z^{\alpha }{}_{\beta \gamma }{}^{\mu }{}_{\nu \sigma }$
{\large . }

\vspace{1cm}

\noindent

\begin{center}
{\large \bigskip \newpage }{\LARGE 11. The System of Macroscopic Gravity
Equations - 3}
\end{center}

\vspace{1cm}

\noindent {\large  $\Rightarrow$  The non-Riemannian Sector:}

{\large \medskip \bigskip }

{\large (2.A) Determine the non-Riemannian curvature tensor }$R^{\alpha
}{}_{\beta \rho \sigma }${\large \ from the Riemannian curvature tensor }$
M^{\alpha }{}_{\beta \rho \sigma }${\large \ and the correlation tensor }$
Q^{\alpha }{}_{\beta \rho \sigma }${\large \ }
\begin{equation}
R^{\alpha }{}_{\beta \rho \sigma }=M^{\alpha }{}_{\beta \rho \sigma
}+Q^{\alpha }{}_{\beta \rho \sigma }  \label{@R=M+Q}
\end{equation}
{\large in terms of metric functions.}

{\large \medskip \bigskip }

{\large (2.B) For the unknown affine deformation tensor }$A^{\alpha
}{}_{\beta \gamma }${\large \ }
\begin{equation}
A^{\alpha }{}_{\beta \gamma }=\overline{\mathcal{F}}^{\alpha }{}_{\beta
\gamma }-\Pi ^{\alpha }{}_{\beta \gamma }
\end{equation}
{\large solve the algebraic equation }
\begin{equation}
A^{\epsilon }{}_{\beta \lbrack \rho }R^{\alpha }{}_{\underline{\epsilon }
\sigma \lambda ]}-A^{\alpha }{}_{\epsilon \lbrack \rho }R^{\epsilon }{}_{
\underline{\beta }\sigma \lambda ]}=0  \label{@AR-AR=0}
\end{equation}
{\large to find some unknown components of the affine deformation tensor }$
A^{\alpha }{}_{\beta \rho }${\large \ in terms of metric functions of the
macroscopic metric tensor }$G_{\alpha \beta }.$

{\large \medskip \bigskip }

{\large (2.C) solve the system of equations for the affine deformation
tensor }$A^{\alpha }{}_{\beta \rho }${\large \ }
\begin{equation}
A^{\alpha }{}_{\beta \lbrack \sigma \parallel \rho ]}-A^{\alpha
}{}_{\epsilon \lbrack \rho }A^{\epsilon }{}_{\underline{\beta }\sigma ]}=-
\frac{1}{2}Q^{\alpha }{}_{\beta \rho \sigma }  \label{@DA=Q}
\end{equation}
{\large for all other components of the affine deformation tensor }$
A^{\alpha }{}_{\beta \rho }${\large .}

{\large \medskip \bigskip }

{\large The steps (2.A)-(2.C) solve completely the non-Riemannian Sector, to
find the affine deformation tensor }$A^{\alpha }{}_{\beta \rho }${\large ,
the non-metric connection }$\Pi ^{\alpha }{}_{\beta \gamma }${\large , the
non-Riemannian curvature tensor }$R^{\alpha }{}_{\beta \rho \sigma }${\large
\ and the non-metricity object }$N_{\alpha \beta \rho }${\large .}

\vspace{1cm}

\noindent {\large  $\Rightarrow$  The Correlation Sector:}

{\large \medskip \bigskip }

{\large (3.A) solve the quadratic algebraic conditions for the connection
correlation tensor }$Z^{\alpha }{}_{\beta \gamma }{}^{\mu }{}_{\nu \sigma }$
{\large \ }
\begin{gather}
Z^{\delta }{}_{\beta \lbrack \gamma }{}^{\theta }{}_{\underline{\kappa }\pi
}Z^{\alpha }{}_{\underline{\delta }\epsilon }{}^{\mu }{}_{\underline{\nu }
\sigma ]}+Z^{\delta }{}_{\beta \lbrack \gamma }{}^{\mu }{}_{\underline{\nu }
\sigma }Z^{\theta }{}_{\underline{\kappa }\pi }{}^{\alpha }{}_{\underline{
\delta }\epsilon ]}+Z^{\alpha }{}_{\beta \lbrack \gamma }{}^{\delta }{}_{
\underline{\nu }\sigma }Z^{\mu }{}_{\underline{\delta }\epsilon }{}^{\theta
}{}_{\underline{\kappa }\pi ]}+  \notag \\
\quad Z^{\alpha }{}_{\beta \lbrack \gamma }{}^{\mu }{}_{\underline{\delta }
\epsilon }Z^{\theta }{}_{\underline{\kappa }\pi }{}^{\delta }{}_{\underline{
\nu }\sigma ]}+Z^{\alpha }{}_{\beta \lbrack \gamma }{}^{\theta }{}_{
\underline{\delta }\epsilon }Z^{\mu }{}_{\underline{\nu }\sigma }{}^{\delta
}{}_{\underline{\kappa }\pi ]}+Z^{\alpha }{}_{\beta \lbrack \gamma
}{}^{\delta }{}_{\underline{\kappa }\pi }Z^{\theta }{}_{\underline{\delta }
\epsilon }{}^{\mu }{}_{\underline{\nu }\sigma ]}=0.  \label{@ZZ=0}
\end{gather}
{\large to find some other independent components of of }$Z^{\alpha
}{}_{\beta \gamma }{}^{\mu }{}_{\nu \sigma }${\large .}

\begin{center}
{\large \bigskip \newpage }{\LARGE 11. The System of Macroscopic Gravity
Equations - 4}
\end{center}

{\large \vspace{1cm}}

\noindent {\large  $\Rightarrow$  The Field Equation Sector:}

{\large \medskip \bigskip }

{\large (4.A) determine the gravitational stress-energy tensor }$T_{\beta
}^{\alpha {\text{(\emph{grav}) }}}${\large \ of macroscopic gravity }
\begin{equation}
(Z^{\alpha }{}_{\mu \nu \beta }-\frac{1}{2}\delta _{\beta }^{\alpha }Q_{\mu
\nu })\bar{g}^{\mu \nu }=-\kappa T_{\beta }^{\alpha {(\text{\emph{grav}) }}}
\label{@RicZ=T(grav)}
\end{equation}
{\large through the components of of }$Z^{\alpha }{}_{(\mu \nu )\beta }$
{\large \ expressed in terms of unknown metric and correlation tensor
functions. An additional assumption regarding the structure of the
gravitational stress-energy tensor }$T_{\beta }^{\alpha {\text{(\emph{grav})}}}$
{\large \ may be necessary.}

{\large \medskip \bigskip }

{\large (4.B) Assume that the averaged inverse metric }$\bar{g}^{\alpha
\beta }${\large \ has the structure }
\begin{equation}
\bar{g}^{\alpha \beta }=G^{\alpha \beta },  \label{@g(-1)b=G(-1)}
\end{equation}
{\large that is, }$\lambda =1${\large \ and the metric reducibility tensor }$
H^{\alpha \beta }=0${\large , which is a reasonable requirement provided the
macroscopic space-time is highly symmetric.}

{\large \medskip \bigskip }

{\large (4.C) Assume that the averaged microscopic stress-energy tensor }$
\langle t_{\beta }^{\alpha \mathrm{(micro)}}\rangle ${\large \ can be taken
as a perfect fluid energy-momentum tensor }

\begin{equation}
\langle \mathbf{t}_{\beta }^{\alpha \mathrm{(micro)}}\rangle =(\rho
+p)u^{\alpha }u_{\beta }+p\delta _{\beta }^{\alpha }  \label{@<t>=pf}
\end{equation}
{\large with the mass density }$\rho ${\large , the pressure }$p${\large \
and the fluid 4-velocity }$u^{\alpha }${\large , }$u^{\alpha }u_{\beta }=-1$
{\large , on the basis of phenomenological considerations with an equation
of state }

\begin{equation}
p=p(\rho ).  \label{@rho(p)}
\end{equation}

{\large \medskip \bigskip }

{\large (4.D) Solve the averaged Einstein's equations }
\begin{equation}
G^{\alpha \epsilon }M_{\epsilon \beta }-\frac{1}{2}\delta _{\beta }^{\alpha
}G^{\mu \nu }M_{\mu \nu }=-\kappa \langle \mathbf{t}_{\beta }^{\alpha
\mathrm{(micro)}}\rangle +(Z^{\alpha }{}_{\mu \nu \beta }-\frac{1}{2}\delta
_{\beta }^{\alpha }Q_{\mu \nu })G^{\mu \nu }  \label{@M=<t>+Z}
\end{equation}
{\large for the unknown metric and correlation functions together with
remaining equations for }$Z^{\alpha }{}_{\beta \gamma }{}^{\mu }{}_{\nu
\sigma }${\large \ and }$A^{\alpha }{}_{\beta \gamma }${\large \ following
from the Riemannian, non-Riemannian and Correlation Sectors.}

\begin{center}
{\large \bigskip \newpage }{\LARGE 12. An Exact Cosmological Solution}
\end{center}

\vspace{1cm}

\noindent {\large  $\Rightarrow$  Consider a flat spatially homogeneous,
isotropic macroscopic space-time given by the Robertson-Walker\ line element
with conformal time }$\eta ${\large \ }
\begin{equation}
ds^{2}=a^{2}(\eta )(-d\eta ^{2}+dx^{2}+dy^{2}+dz^{2})  \label{rw-flat}
\end{equation}
{\large where }$d\eta =a^{-1}(t)dt${\large \ with a cosmological
(coordinate) time }$t${\large \ and }$a^{2}(t)${\large \ is an unknown
function of the scale factor. }

\vspace{1cm}

\noindent {\large  $\Rightarrow$  Make an ansatz about the functional form
of the components of the correlation tensor }$Z^{\alpha }{}_{\beta \gamma
}{}^{\mu }{}_{\nu \sigma }${\large \ on the basis of symmetries and physical
conditions of the macroscopic geometry (\ref{rw-flat}). The simplest
possible condition on }$Z^{\alpha }{}_{\beta \gamma }{}^{\mu }{}_{\nu \sigma
}${\large \ here which appears to be compatible with the structure of
macroscopic space-time (\ref{rw-flat}) is to require all of its components
to be constant }
\begin{equation}
Z^{\alpha }{}_{\beta \gamma }{}^{\mu }{}_{\nu \sigma }=\mathrm{const}.
\label{Z6=const}
\end{equation}
{\large From the physical point of view this condition means that we assume
that the macroscopic gravitational correlations do not change in time and
space. }

\vspace{1cm}

\noindent {\large  $\Rightarrow$  The Macroscopic Gravity equations can be
solved now to show that there is finally only one remaining independent
component }$Z^{3}{}_{23}{}^{3}{}_{32}$ {\large of the connection correlation
tensor }$Z^{\alpha }{}_{\beta \gamma }{}^{\mu }{}_{\nu \sigma }${\large \
which is determined through an integration constant. Upon denoting constant }
$12Z^{3}{}_{23}{}^{3}{}_{32}=-\varepsilon ${\large \ the macroscopic
gravitational stress-energy tensor has been shown to take the form }
\begin{equation}
\kappa T_{\alpha \beta }^{{(\text{\emph{grav}})}}{}=\kappa G_{\alpha \epsilon
}T_{\beta }^{\epsilon {(\text{\emph{grav}})}}{}=
\begin{bmatrix}
\varepsilon & 0 & 0 & 0 \\
0 & -\frac{1}{3}\varepsilon & 0 & 0 \\
0 & 0 & -\frac{1}{3}\varepsilon & 0 \\
0 & 0 & 0 & -\frac{1}{3}\varepsilon
\end{bmatrix}
\label{Tgrav[2,2,3]real-cov}
\end{equation}
{\large where }$\varepsilon /\kappa a^{2}=\rho _{\text{\emph{grav}}}${\large \ is
the macroscopic gravitational correlation energy density and }$-\varepsilon
/3\kappa a^{2}=p_{\text{\emph{grav}}}${\large \ is the isotropic pressure of
macroscopic gravitational correlation field. Thus, the equation of state for
the macroscopic gravitational correlation field is\medskip \bigskip }
\begin{equation}
\fbox{$p_{\text{\emph{grav}}}=p_{\text{\emph{grav}}}(\rho _{\text{\emph{grav}}})=-\frac{1}{3}\rho
_{\text{\emph{grav}}}$}  \label{MG-EqSt}
\end{equation}

\begin{center}
{\large \bigskip \newpage }{\LARGE 12. An Exact Cosmological Solution - 2}
\end{center}

\vspace{1cm}

\noindent {\large  $\Rightarrow$  The trace of }$T_{\beta }^{\epsilon
{\text{(\emph{grav}})}}{}$ is {\large \ }
\begin{equation}
T_{\epsilon }^{\epsilon {\text{(\emph{grav}})}}{}=-\frac{2\varepsilon }{\kappa a^{2}}
=-2\rho _{\text{\emph{grav}}},\quad T_{\epsilon }^{\epsilon {\text{(\emph{grav}})}}<0~
\mathrm{if~}\rho _{\text{\emph{grav}}}>0,  \label{Tr_Tgrav[2,2,3]real}
\end{equation}
{\large that means from the physical point of view that the macroscopic
energy is the bounding energy of the Universe. It acts like an tension in an
elastic medium to keep it intact. An amount of work should be performed and
an amount of energy should be used to change its state, size and shape. A
negative pressure has the similar physical meaning. The bounding energy
density is decreasing with an increasing scale factor. }

{\large \vspace{1cm}}

\noindent {\large  $\Rightarrow$  A macroscopic distribution of the
cosmological matter through the averaged microscopic stress-energy tensor }$
\langle t_{\beta }^{\alpha \mathrm{(micro)}}\rangle ${\large \ is taken as a
perfect fluid energy-momentum tensor (\ref{@<t>=pf}), (\ref{@rho(p)}).}
\begin{equation}
\langle \mathbf{t}_{\alpha \beta }^{\mathrm{(micro)}}\rangle =G_{\alpha
\epsilon }\langle \mathbf{t}_{\beta }^{\epsilon \mathrm{(micro)}}\rangle =
\begin{bmatrix}
\rho a^{2} & 0 & 0 & 0 \\
0 & pa^{2} & 0 & 0 \\
0 & 0 & pa^{2} & 0 \\
0 & 0 & 0 & pa^{2}
\end{bmatrix}
.  \label{<t>=pf}
\end{equation}

\vspace{1cm}

\noindent {\large  $\Rightarrow$  After transformation from the conformal
time }$\eta ${\large \ to the cosmological time }$t${\large \ the averaged
Einstein equations (\ref{@M=<t>+Z}) read }
\begin{gather}
\left( \frac{\dot{a}}{a}\right) ^{2}=\frac{\kappa \rho }{3}+\frac{
\varepsilon }{3a^{2}},  \label{macro-law1} \\
2\frac{\ddot{a}}{a}+\left( \frac{\dot{a}}{a}\right) ^{2}=-\kappa p+\frac{
\varepsilon }{3a^{2}},  \label{macro-law2}
\end{gather}
{\large or in terms of }$\rho _{\text{\emph{grav}}}${\large \ and }$p_{\text{\emph{grav}}}$
{\large \ }
\begin{gather}
\left( \frac{\dot{a}}{a}\right) ^{2}=\frac{\kappa }{3}\left( \rho +\rho _{
{\text{\emph{grav}}}}\right) ,  \label{law1} \\
2\frac{\ddot{a}}{a}+\left( \frac{\dot{a}}{a}\right) ^{2}=-\kappa \left( p+p_{
{\text{\emph{grav}}}}\right) .  \label{law2}
\end{gather}
{\large with the equations of state }$p=p(\rho )$ {\large (\ref{@rho(p)})
and }$p_{\text{\emph{grav}}}=-\frac{1}{3}\rho _{\text{\emph{grav}}}${\large \
(\ref{MG-EqSt}). }

\begin{center}
{\large \bigskip \newpage }{\LARGE 12. An Exact Cosmological Solution - 3}
\end{center}

\vspace{1cm}

\noindent {\large  $\Rightarrow$  They look similar to Einstein's equations
of General Relativity for either a closed or an open spatially homogeneous,
isotropic FLRW space-time, but they do have different mathematical and
physical, and therefore, cosmological content since }
\begin{equation}
\frac{\varepsilon }{3}=\frac{\kappa \rho _{\text{\emph{grav}}}a^{2}}{3}\neq -k
\label{sp-curv-not-k}
\end{equation}
{\large in general. }

\vspace{1cm}

\noindent {\large  $\Rightarrow$  The macroscopic (averaged) Einstein's
equations for a flat spatially homogeneous, isotropic macroscopic space-time
have macroscopic gravitational correlation terms of the form of a spatial
curvature term }
\begin{equation}
\frac{\varepsilon }{3a^{2}}=\frac{\kappa \rho _{\text{\emph{grav}}}}{3}.
\label{sp-curv}
\end{equation}
{\large Thus, the theory of Macroscopic Gravity predicts that constant
macroscopic gravitational correlation tensor }
\begin{equation}
Z^{\alpha }{}_{\beta \gamma }{}^{\mu }{}_{\nu \sigma }=\mathrm{const}.
\label{Z6=c2}
\end{equation}
{\large for a flat spatially homogeneous, isotropic macroscopic space-time }
\begin{equation}
ds^{2}=a^{2}(\eta )(-d\eta ^{2}+dx^{2}+dy^{2}+dz^{2})  \label{SHI}
\end{equation}
{\large takes the form of }\textsc{\ \medskip \bigskip }

\bigskip

\begin{center}
\fbox{\textbf{a} \textbf{dark spatial curvature term}}
\end{center}

\textsc{\ \medskip \bigskip }

$\bigstar $ \textsc{\ it interacts only gravitationally with the macroscopic
gravitational field \medskip \bigskip }

$\bigstar $ \textsc{\ it does not interact directly with the energy-momentum
tensor of matter \medskip \bigskip }

$\bigstar $ \textsc{\ it exhibits a negative pressure }$p_{\text{\emph{grav}}}=-
\frac{1}{3}\rho _{\text{\emph{grav}}}$ \textsc{which tends to accelerate the
Universe when }$\rho _{\text{\emph{grav}}}>0$\textsc{.}

\bigskip

\bigskip

\begin{center}
{\large \bigskip \newpage }{\LARGE 12. An Exact Cosmological Solution - 4}
\end{center}

\vspace{1cm}

\noindent {\large  $\Rightarrow$  Only if one requires }$
12Z^{3}{}_{23}{}^{3}{}_{32}=-\varepsilon ${\large \ to be }
\begin{equation}
\varepsilon =-3k  \label{k=-1}
\end{equation}
{\large the macroscopic (averaged) Einstein's equations become exactly
Einstein's equations of General Relativity for either a closed or an open
spatially homogeneous, isotropic space-time for the macroscopic geometry of
a flat spatially homogeneous, isotropic space-time. }

\vspace{1cm}

\noindent {\large  $\Rightarrow$  This exact solution of the Macroscopic
Gravity equations exhibits a very non-trivial phenomenon from the point of
view of the general-relativistic cosmology: }

\vspace{1cm}

\begin{center}
\textbf{the macroscopic (averaged) cosmological evolution in a flat Universe}

\textbf{is governed by the dynamical evolution equations for either a closed
}

\textbf{\ or an open Universe depending on the sign of the macroscopic energy
}

\textbf{\ density }$\rho _{\text{\emph{grav}}}$\textbf{\ with a dark spatial
curvature term}{\large \ }$\kappa \rho _{\text{\emph{grav}}}/3$
\end{center}

\vspace{1cm}

\noindent {\large  $\Rightarrow$  From the observational point of view such
a cosmological model gives a new paradigm to reconsider the standard
cosmological interpretation and treatment of the observational data. }

{\large Indeed, this macroscopic cosmological model has the Riemannian
geometry of a flat homogeneous, isotropic space-time. Therefore, all
measurements and data are to be considered and designed for this geometry.
The dynamical interpretation of the obtained data should be considered and
treated for the cosmological evolution of either a closed or an open
spatially homogeneous, isotropic Riemannian space-time.}

\begin{center}
\bigskip \newpage
\end{center}

\end{document}